\documentclass[prl,twocolumn,superscriptaddress,showpacs]{revtex4}
\usepackage{graphicx}% Include figure files
\usepackage{epsfig}% Include figure files
\usepackage{amssymb,amsmath,amsfonts,hyperref}
\usepackage{wasysym}
\usepackage{latexsym}
\usepackage{eucal}
\usepackage{verbatim}
\usepackage[normalem]{ulem}

\usepackage{color}

\newcommand{\be}{\begin{eqnarray}}
\newcommand{\ee}{\end{eqnarray}}

\begin{document}
%\preprint{APS/123-QED}
%draft
%\twocolumn[\hsize\textwidth\columnwidth\hsize\csname @twocolumnfalse\endcsname
\title{Melting of three-sublattice order in easy-axis antiferromagnets on triangular and Kagome lattices.}
\author{Kedar Damle}
\affiliation{\small{Tata Institute of Fundamental Research, 1 Homi Bhabha Road, Mumbai 400005, India}}
\begin{abstract}
When the constituent spins have an energetic preference to lie along an easy-axis, triangular and Kagome lattice antiferromagnets often develop long-range order that
distinguishes the three sublattices of the underlying triangular Bravais lattice. In zero magnetic field, this three-sublattice order melts
 {\em either} in a two-step manner, {\em i.e.} via an intermediate phase
with power-law three-sublattice order controlled by a temperature dependent exponent $\eta(T) \in (\frac{1}{9},\frac{1}{4})$,
{\em or} via a transition in the three-state Potts universality class. Here, I predict that the uniform susceptibility to a small easy-axis field $B$ diverges
as $\chi(B) \sim |B|^{-\frac{4 - 18 \eta}{4-9\eta}}$  in a large part of the intermediate power-law ordered phase (corresponding to $\eta(T) \in (\frac{1}{9},\frac{2}{9})$), providing an easy-to-measure thermodynamic signature of two-step melting. I also show that these two melting scenarios can be generically connected via an intervening multicritical point, and obtain numerical estimates of multicritical exponents.

\end{abstract}

\vskip2pc
%\date{\today}
\maketitle
In frustrated antiferromagnets~\cite{frustratedreview1,fragnetsreview1},  magnetic ions (spins) form a lattice whose geometry
causes the dominant antiferromagnetic interactions between neighbours to compete with each other. This allows weaker further-neighbour interactions or quantum fluctuations to select complex patterns of spin order at low temperature.
Models of frustrated easy-axis antiferromagnets~\cite{Sen_Wang_Damle_Moessner,Sen_Damle_Vishwanath}, in which spins can lower energy by orienting
along a fixed axis, provide interesting examples of this behaviour.

Such models are also relevant in other experimental contexts. 
For instance, the low-temperature behaviour
of monolayers of adsorbed gases on substrates with triangular symmetry~\cite{Bretz_etal,Bretz,Horn_Birgeneau_Heiney_Hammonds,Vilches,Suter_Colella_Gangwar,Feng_Chan,Wiechert} has been
modeled~\cite{Landau} in terms of a triangular lattice of Ising spins $\sigma^z_{\vec{R}}=\pm1$ ($\hat{z}$  components of spin-half moments $\vec{S}_{\vec{R}}=\vec{\sigma}_{\vec{R}}/2$) with antiferromagnetic Ising interactions $J\sigma^z_{\vec{R}} \sigma^z_{\vec{R}'}$ between nearest-neighbours~\cite{Wannier,Stephenson} and weak ferromagnetic Ising interactions between further-neighbours. 
More recently, the magnetic
properties of honeycomb networks~\cite{Tanaka_etal,Qi_etal,Ladak_etal1,Ladak_etal2} of magnetic wires (dubbed artificial Kagome-ice) have been analyzed~\cite{Moller_Moessner,Chern_etal,Chern_Tchernyshyov} in terms of a similar Ising model on the
Kagome lattice~\cite{Kano_Naya}. In
both examples, further-neighbour couplings cause the Ising spins to 
develop ferrimagnetic three-sublattice order at low temperature, {\em i.e.}, freeze into a pattern which distinguishes the three sublattices of the underlying triangular Bravais lattice and gives rise to a small net moment along the easy axis.

Several other easy-axis spin systems on triangular and Kagome lattices exhibit ferrimagnetic three-sublattice order~\cite{Wolf_Schotte,Takagi_Mekata,Wills_Ballou_Lacroix,Nienhuis_Hilhorst_Blote,Zeng_Henley,Melko_etal,Heidarian_Damle,Wessel_Troyer,Auerbach_Murthy,Boninsegni,Sen_etal,Damle_Senthil}, or closely related antiferromagnetic  (no
net easy-axis moment) three-sublattice order~\cite{Isakov_Moessner}. 
In zero field ($B=0$) along the easy-axis, a Ginzburg-Landau theory~\cite{Domany_Schick_Walker_Griffiths,Domany_Schick,Alexander} for the three-sublattice order parameter predicts that this ordering transition is described by a six-fold anisotropic effective model of ferromagnetically coupled $XY$ spins~\cite{Jose}, or, equivalently, by a generalized six-state clock model~\cite{Cardy,Dorey_etal,Tobochnik,Challa_Landau}. Rather unusually, such six-state clock
models have multiple generic possibilities for continuous transitions: Order
is lost {\em either} via
a two-step melting transition, with an intermediate phase characterized by power-law order~\cite{Jose}, {\em or} via a sequence of  two distinct transitions, one of which is in the three-state Potts universality class and the other in the Ising universality class~\cite{Cardy,Dorey_etal}.
Perhaps motivated by this, the melting of three-sublattice order has been
studied in a variety of triangular and Kagome lattice systems for over three decades now. In some examples~\cite{Landau,Nienhuis_Hilhorst_Blote,Wolf_Schotte,Takagi_Mekata,Wills_Ballou_Lacroix,Isakov_Moessner}, three-sublattice order is known to melt in a two-step manner, via a sizeable intermediate phase with power-law three-sublattice order controlled by a temperature-dependent exponent $\eta(T) \in  (\frac{1}{9},\frac{1}{4})$. In other examples with {\em ferrimagnetic} three-sublattice order, this order is lost via a three-state Potts transition, while residual {\em ferromagnetism} is lost via an Ising transition~\cite{Moller_Moessner,Chern_etal,Chern_Tchernyshyov}.

In this Letter, I analyze the melting of three-sublattice order in easy-axis antiferromagnets
on triangular and Kagome lattices using a new coarse-grained description that explicitly keeps track
of the uniform easy-axis magnetization mode whose fluctuations are coupled to fluctuations of the three-sublattice order parameter.
Using this description, which goes beyond the standard Ginzburg-Landau theory, I demonstrate that these two very different melting processes can be generically connected via an
intervening multicritical point ${\mathcal M}$ (Fig.~\ref{Fig2}) with central charge~\cite{CardyBook} $c_{{\mathcal M}} \in (1,\frac{3}{2})$. Although the generalized six-state clock model correctly captures other generic ways~\cite{Cardy,Dorey_etal} in which these two very different melting processes can be separated from each other in the phase diagram of such three-sublattice ordered systems, it fails to account for the existence of ${\cal M}$. This underscores the importance of treating the uniform magnetization mode on the same footing as the three-sublattice order parameter.

I obtain numerical estimates of multicritical exponents, and argue that such multicritical melting may be experimentally accessible in artificial Kagome-ice systems
if the strengths of nearest and next-nearest exchange interactions can be increased relative to the long-range dipolar interactions. 
Additionally, for $\eta(T) \in (\frac{1}{9},\frac{2}{9})$ in the power-law ordered phase associated with two-step melting, I show that the {\em uniform susceptibility} to a small easy-axis field $B$ diverges as $\chi(B) \sim |B|^{-\frac{4 - 18 \eta}{4-9\eta}}$. I also argue that this easy-to-measure thermodynamic signature is of potential experimental relevance in the context of
three-sublattice ordering of nearly-half-filled monolayers of adsorbed gases on triangular substrates, and in the context of experimental realizations of three-sublattice order
in $S=1$ Heisenberg antiferromagnets with strong single-ion anisotropy on the triangular lattice.

\begin{figure}
{\includegraphics[width=\hsize]{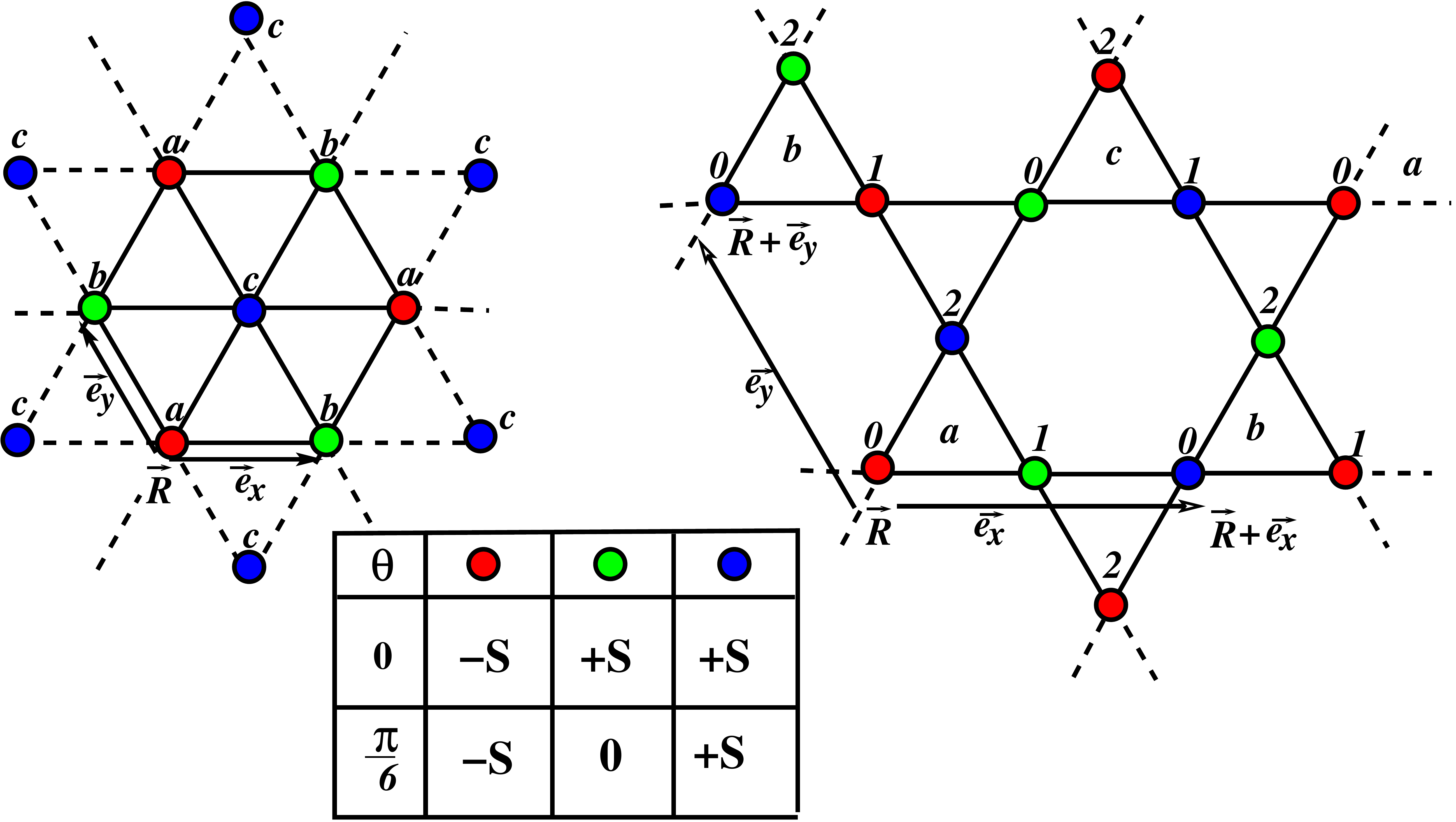}}
\caption{Color-coded symbols on sites give
the value of $\langle S^z_{\vec{r}} \rangle$ in the presence of ferrimagnetic
($\theta = 0$) or antiferromagnetic ($\theta=\frac{\pi}{6}$) three-sublattice order  in spin-$S$ triangular [Kagome] lattice easy-axis antiferromagnets. These ordering patterns distinguish between the three sublattices of the underlying Bravais lattice of sites [up-pointing triangles]. }
\label{Fig1}
\end{figure}

\begin{figure}
{\includegraphics[width=\hsize]{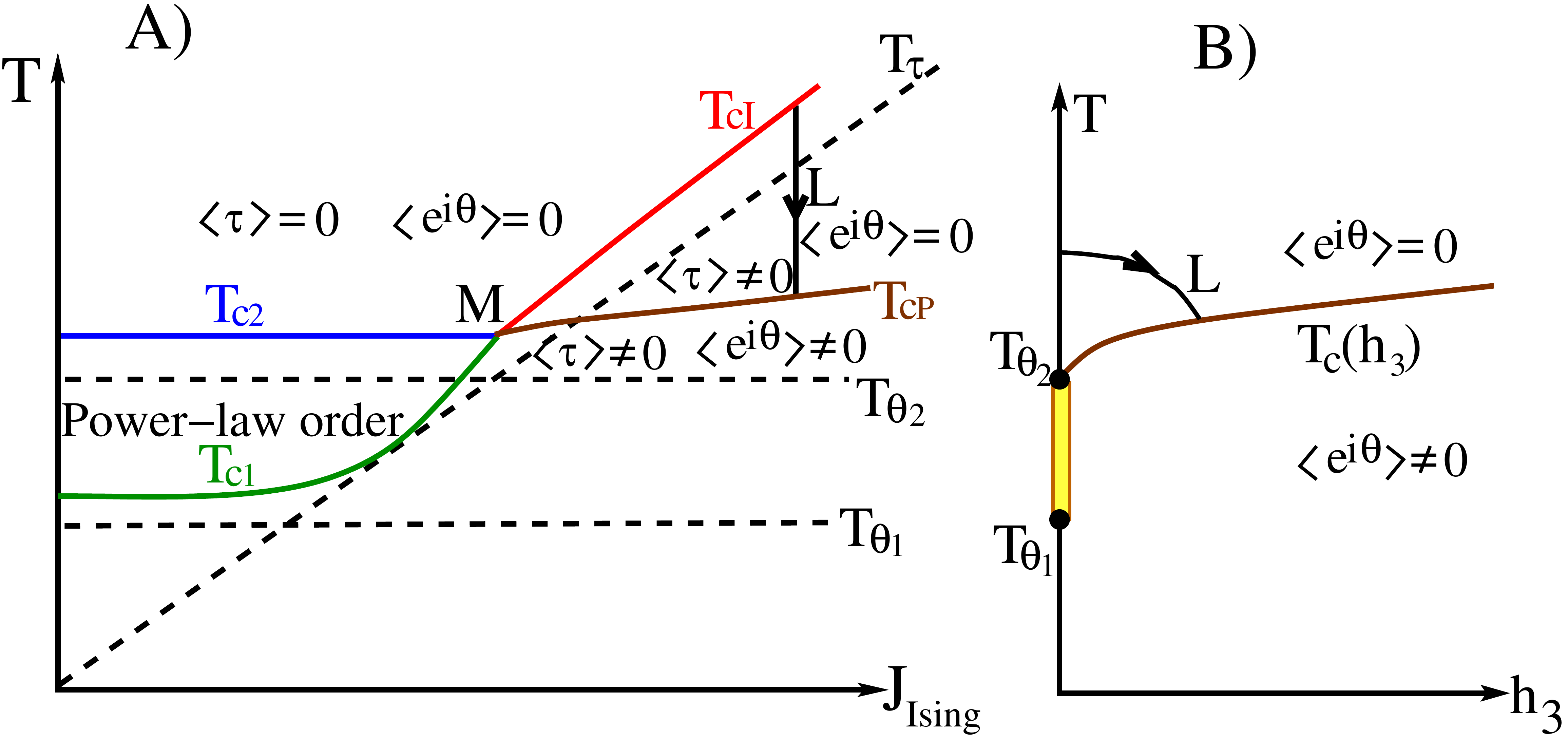}}
\caption{(Color online) A) Predicted structure of the
$T$-$J_{\rm Ising}$ phase
diagram of $H_{\rm eff}$ for $h=0$ and fixed $J_{\rm xy}$ and $J_{\theta \tau}$. Phase boundaries of $H_{\rm eff}$ are depicted by colour-coded solid lines, while those of $H_{\rm Ising}$ and $H_{\rm xy}$ are displayed as dashed lines. 
B) Known $T$-$h_3$ phase diagram of $H_{\rm xy} + h_3 \sum_{\vec{r}}\cos(3\theta_{\vec{r}})$
showing the three-state Potts line $T_c(h_3)$. Path $L$ in A) maps to the eponymous path in B).}
\label{Fig2}
\end{figure}

{\bf Order parameters and coarse-graining:} I use the convention of Fig.~(\ref{Fig1})  for labeling the sites [unit-cells] $\vec{R}=m\hat{e}_x+n\hat{e}_y$ of
the triangular [Kagome] lattice, and for labeling the three basis sites $\alpha=0,1,2$ in each unit-cell of the Kagome lattice. With this convention, the complex three-sublattice order parameter $\psi \equiv |\psi|e^{i \theta}$ and the ferromagnetic order parameter $M^z$
are defined as: $\psi = -\sum_{\vec{R}} e^{i \frac{2 \pi}{3}(m+n)} S^z_{\vec{R}}$ and  $M^z = \sum_{\vec{R}} S^z_{\vec{R}}$ on the triangular
lattice, while $\psi = -\sum_{\vec{R},\alpha}e^{i \frac{2 \pi}{3}(m+n-\alpha)} S^z_{\vec{R},\alpha} $ and $M^z = \sum_{\vec{R},\alpha}S^z_{\vec{R},\alpha} $ on the Kagome lattice. Our coarse-grained
description will be written in terms of an effective Hamiltonian defined on a lattice whose sites $\vec{r}$ represent clusters of spins of the original triangular or Kagome magnet. In this description, each cluster is characterized by an Ising variable $\tau_{\vec{r}} = \pm 1$ representing the direction of the local easy-axis magnetization $M^{z}_{\rm cluster}$, and by an angle $\theta_{\vec{r}}$ that represents the phase of the local
three-sublattice order parameter $\psi_{\rm cluster}$. Comparison with long-wavelength properties of specific microscopic models is facilitated by choosing clusters that themselves form a coarse-grained triangular lattice, since this preserves the symmetries of
the underlying triangular Bravais lattice in both triangular and Kagome lattice systems.

{\bf Ginzburg-Landau theory:} Let us begin by summarizing in this language the standard Ginzburg-Landau theory for three-sublattice ordering~\cite{Domany_Schick_Walker_Griffiths,Domany_Schick,Alexander}: Transformation
properties of $\psi$ under global spin-flip and lattice symmetry operations
fix the form of the effective Hamiltonian $H_{\rm xy}$ for $\theta_{\vec{r}}$. Leaving out
certain chiral perturbations~\cite{Huse_Fisher_PRB,Huse_Fisher,Kardar_Berker} that are not expected to be
relevant~\cite{Huse} for the transitions of the lattice magnets studied here,
$H_{\rm xy}$ may be written as
\begin{eqnarray}
H_{\rm xy} &=& -J_{\rm  xy} \sum_{\langle \vec{r} \vec{r}' \rangle} \cos(\theta_{\vec{r}} - \theta_{\vec{r}'})
- h_6\sum_{\vec{r}}\cos(6\theta_{\vec{r}}) \; .
\end{eqnarray}
where $\langle \vec{r} \vec{r}' \rangle$ are nearest-neighbour links of our coarse-grained triangular lattice.
The effective stiffness $J_{\rm xy} > 0 $ (encoding the energetic preference for three-sublattice order) and  the six-fold anisotropy $h_6$, whose sign selects between ferrimagnetic
three-sublattice order (with  $\theta_m=2\pi m/6$) and antiferromagnetic three-sublattice
order ($\theta_m=(2m+1)\pi/6$), are both set by quantum fluctuations and subdominant further-neighbour couplings in the microscopic Hamiltonian.
In this approach, the relative values of $J_{\rm xy}$ and its higher harmonics
$J_{\rm xy}^{(p)}$ (coefficients of $-\cos(p\theta_{\vec{r}} - p\theta_{\vec{r}'})$ for $p=2,3$)
determine the nature of the melting process. These higher harmonics are omitted from $H_{\rm xy}$ displayed above since they are not crucial for our subsequent discussion.

{\bf New effective Hamiltonian:} Next, I note that this standard Ginzburg-Landau description does 
not take into account the uniform magnetization mode whose fluctuations are
coupled in a crucial way to fluctuations of the three-sublattice order parameter. This
key observation leads me to a new coarse-grained effective model:
\begin{eqnarray}
H_{\rm  eff} &=& H_{\rm  xy} + H_{\rm  Ising} -J_{\theta \tau} \sum_{\vec{r}} \tau_{\vec{r}} \cos(3\theta_{\vec{r}}) \; , \nonumber \\
&& \\
{\rm where} \; \; \; H_{\rm Ising} &=&-J_{\rm  Ising} \sum_{\langle \vec{r} \vec{r}' \rangle} \tau_{\vec{r}} \tau_{\vec{r}'} -h\sum_{\vec{r}} \tau_{\vec{r}} \; , \nonumber
\end{eqnarray}
with  $h \propto B$.
To understand the rationale for the form of this effective Hamiltonian, it is useful to
first note that $H_{\rm eff}$ has the same $S_3 \times Z_2$ symmetry as $H_{\rm xy}$, and reduces, in the double limit $h_6,J_{\theta \tau} \rightarrow \infty$, to a generalized six-state clock model studied earlier~\cite{Cardy,Dorey_etal}. However, the space of states at each site of $H_{\rm eff}$ is enlarged by the presence of $\tau_{\vec{r}}$ to
correctly account for the fact that the direction of $M^z_{\rm cluster}$ is correlated with
the phase of $\psi_{\rm cluster}$, but not completely tied to it. The microscopic origin
of various terms can now be understood as follows:
 $J_{\rm Ising} > 0 $ encodes the effect of subleading ferromagnetic interactions of the microscopic magnet, which tend to favour ferrimagnetic three-sublattice order. If $h_6 > 0$, it is likely to be
accompanied by a sizeable positive value of  $J_{\rm Ising}$ in $H_{\rm eff}$ (since ferrimagnetic three-sublattice order corresponds to $h_6>0$ in $H_{\rm xy}$). Conversely, negative $h_6$, favoured by quantum-fluctuations in some
examples~\cite{Isakov_Moessner}, is likely to be accompanied by negligibly small $J_{\rm  Ising}$.
The coupling $J_{\theta \tau}>0$ correctly captures the
fact that the values $\theta=0, 2\pi/3, 4\pi/3$ ($\pi/3, \pi, 5\pi/3$), characteristic of ferrimagnetic three-sublattice order, are associated with a positive (negative) easy-axis magnetization,
while the phase choices $\theta=(2m+1)\pi/6$, characteristic of antiferromagnetic three-sublattice order, are not associated with any net easy-axis magnetization (Fig.~\ref{Fig1}).

{\bf Phase-diagram of $H_{\rm eff}$:} To deduce the structure of the $h=0$ phase
diagram of $H_{\rm eff}$ (Fig~(\ref{Fig2} A)) in the $T$-$J_{\rm Ising}$  plane (with $J_{\rm xy}=1$) for fixed ${\mathcal O}(1)$ values of $J_{\theta \tau}$ and $h_6$, 
I start with the known phase diagrams of
$H_{\rm  xy}$ and $H_{\rm  Ising}$, and analyze the effects of a non-zero $J_{\theta \tau}$.
To this end, recall that
$H_{\rm Ising}$ develops long-range order in $\tau$ for $T < T_{\tau}$, with long-distance properties of the critical point at  $T_{\tau}$ described by a fixed-point free-energy functional $F_{1/2} = \int d^2x {\cal F}_{1/2}$, with central charge $c=1/2$.
Similarly, $H_{\rm xy}$  develops six-fold symmetry-breaking long-range order in $\theta$ for $T<T_{\theta 1}$, which melts via an intermediate phase with power-law correlations: $\langle e^{i(\theta(\vec{r}) - \theta(0))}\rangle \sim 1/|\vec{r}|^{\eta(T)}$ with $\eta(T) \in (\frac{1}{9}, \frac{1}{4})$ for temperatures $T \in (T_{\theta 1},T_{\theta 2})$~\cite{Tobochnik,Challa_Landau,Rastelli_Regina_Tasi}.
Long-wavelength properties of this power-law ordered phase are controlled, in renormalization group (RG) language, by a $c=1$ line of fixed points~\cite{Jose}, with effective free-energy $F_{\rm KT} = \int d^2r {\cal F}_{\rm KT}$, where 
\begin{equation}
{\cal F}_{\rm KT}/T = \frac{1}{4 \pi g}(\nabla \theta)^2 
\label{KTfixedline}
\end{equation}
with $g(T) \in (\frac{1}{9},\frac{1}{4})$
corresponding to $T \in (T_{\theta 1},T_{\theta 2})$. This fixed-line has power-law correlations $\langle e^{i(\theta(\vec{r})-\theta(0))}\rangle \sim 1/r^{\eta(g)}$ with $\eta(g) = g$, which render the six-fold symmetry-breaking perturbation $h_6\cos(6\theta_{\vec{r}})$ irrelevant for $g>1/9$,
and vortices in $\theta$ irrelevant for $g<1/4$~\cite{Jose}. However, the three-fold symmetric perturbation $h_3\cos(3\theta_{\vec{r}})$ is relevant everywhere on this fixed line~\cite{Jose}, implying that long-range order sets in
at infinitesimal $h_3$ when $T<T_{\theta 2}$. In contrast, for fixed $T>T_{\theta 2}$,
long-range order sets in via a three-state Potts transition~\cite{Jose}(Fig.~\ref{Fig2} B) only when a threshold $h_{3c}(T)$ is crossed; this defines a three-state Potts critical line $T_c(h_3)$  in the $(T,h_3)$ phase diagram of $H_{\rm xy}$ (Fig.~(\ref{Fig2} B)).
Therefore, our analysis splits naturally
into two cases, $T_{\tau} \lesssim T_{\theta 2}$, and $T_{\tau} \gtrsim T_{\theta 2}$, and relies crucially on the observation that long-range order of $\theta$ in $H_{\rm xy}$ leads to an external magnetic field of effective strength $h_{\rm eff} \equiv J_{\theta \tau} \langle \cos(3\theta) \rangle$ acting on $\tau$ in $H_{\rm Ising}$, while long-range order of $\tau$ in $H_{\rm Ising}$
perturbs $H_{\rm xy}$  by a three-fold symmetric term $\sum_{\vec{r}} h_{3{\rm eff}} \cos(3 \theta_{\vec{r}})$, with $h_{3{\rm eff}} \equiv  J_{\theta \tau} \langle \tau \rangle$.

$\mathbf{T_{\tau} \lesssim T_{\theta 2}}${\bf:} If $H_{\rm Ising}$
is in a short-range correlated paramagnetic phase in the entire temperature
range $(T_{\theta 1}, T_{\theta 2})$, {\em i.e.} if $T_{\tau} \lesssim T_{\theta 1}$, a non-zero $J_{\theta \tau}$ only renormalizes
the value of $g(T)$ that controls the power-law correlators of $\theta$ in this regime. And when the
temperature is lowered below $T_{\theta 1}$, long-range order of  $\theta$ in $H_{\rm xy}$ gives rise to an effective field $h_{\rm eff}  \equiv J_{\theta \tau} \langle \cos(3\theta) \rangle$ in $H_{\rm Ising}$, converting the Ising transition at $T_{\tau}$ to a smooth crossover.
On the other hand, if $T_{\theta 1} \lesssim T_{\tau}$, long-range order of $\tau$ below $T_{\tau}$ leads to a {\em three-fold} symmetric perturbation $h_{3 {\rm eff}}$ of $H_{\rm xy}$, which immediately causes $H_{\rm xy}$ to develop long-range order in $\theta$ (Fig.~\ref{Fig2} B).

Thus, when $T_{\tau} \lesssim T_{\theta 2}$, $H_{\rm eff}$ is expected to display
a six-fold symmetry-breaking long-range ordered state for $T < T_{c 1}$, which
undergoes a two-step melting transition via an intermediate power-law ordered phase
(corresponding to $T \in (T_{c1} , T_{c2})$) with an exponent $\eta(T)$ that increases from $\eta(T_{c 1}) =\frac{1}{9}$ to $\eta(T_{c 2}) = \frac{1}{4}$. The value of $T_{c1}$ is
set  (with deviations of order $J_{\theta \tau}$)
by the larger of $T_{\theta 1}$ and $T_{\tau}$, while that of $T_{c2}$ is approximately set by $T_{\theta 2}$.

\begin{figure}
{\includegraphics[width=\hsize]{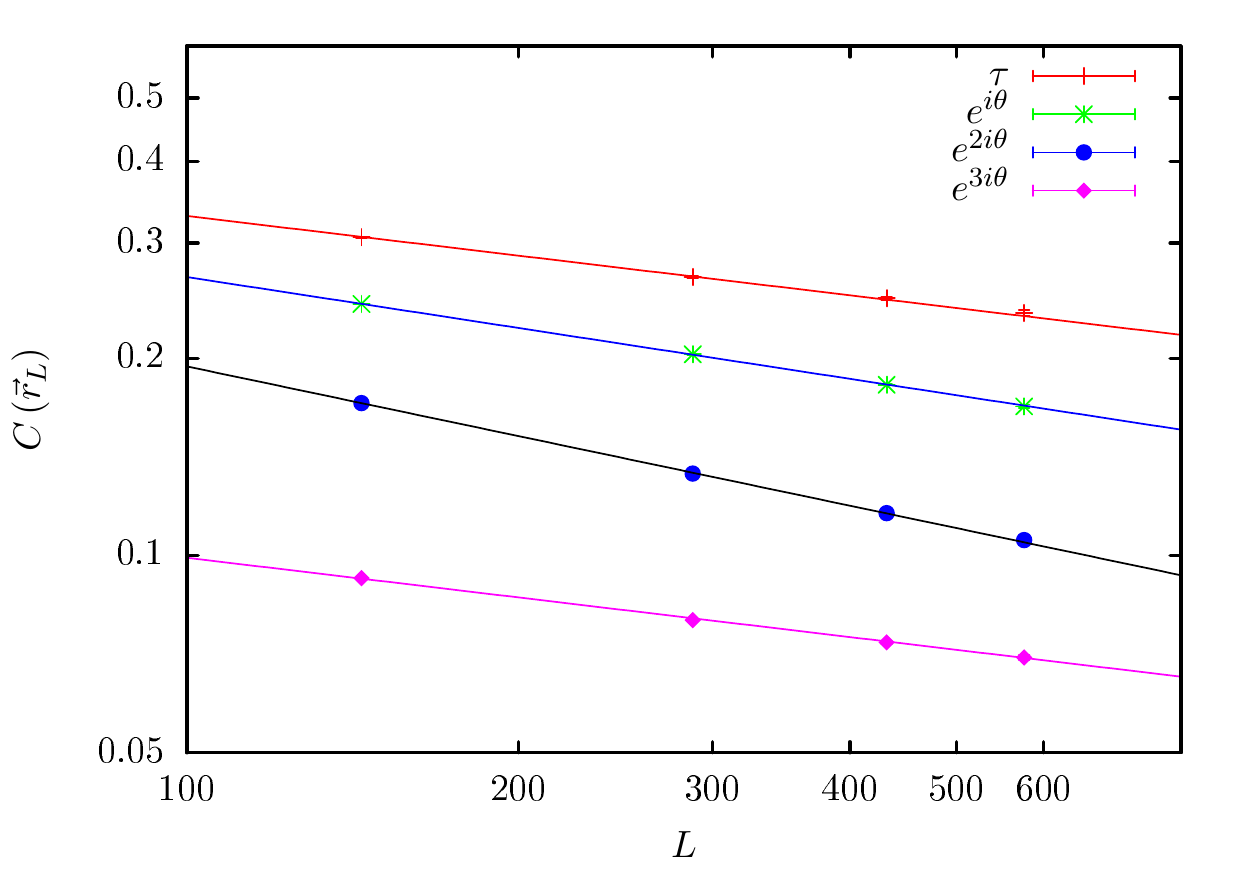}}
\caption{$L$ dependence
of $C_{\tau}(\vec{r}_L)$ and $C_{p\theta}(\vec{r}_L)$ (p=1,2,3) at separation $\vec{r}_L=\hat{e}_x \frac{L}{3}$ on periodic
$L \times L$ triangular lattices, evaluated at the estimated location $[f_{\rm xy}^{{\mathcal M}},f_{I}^{{\mathcal M}}] = [1.5570, 1.0061]$ of the multicritical point of $H_{{\rm eff}}$ with $J_{\rm xy} = h_6 = 1.0$, $J_{\theta \tau}=0.25$ (notation as in text). Lines denote fits to $1/L^{\eta_\tau}$ and $1/L^{\eta_{p\theta}}$ respectively,  using $\eta_{3\theta}=\eta_{\tau}=0.201$, $\eta_{\theta} = 0.258$, and $\eta_{2\theta} = 0.353$. $C_{2\theta}$ [$C_{3\theta}$] is rescaled by a factor of $7$ [factor of $10$] for clarity.}
\label{Fig3}
\end{figure}

{\bf Power-law ordered phase:} Long-wavelength properties of $H_{\rm eff}$ in this power-law ordered intermediate phase can be described quite generally (for either sign of $h_6$ ) 
by an effective free-energy density
\begin{equation}
{\cal F}_{\tau {\rm KT}}/T = {\cal F}_{\rm KT}/T + c_{\theta \tau} \tau_{\vec{r}} \cos(3\theta_{\vec{r}}) \; .
\label{TauKTfixedline}
\end{equation}
Although a nonzero $c_{\theta \tau}$ leads, upon tracing over $\tau$, to the six-fold term
$\cos(6\theta_{\vec{r}})$ which is irrelevant all along the fixed-line parametrized by $c_{\theta \tau} = 0$ and $g(T) \in (\frac{1}{9}, \frac{1}{4})$
[as in in Eqn.~\ref{KTfixedline}], I choose to retain a bare $c_{\tau \theta} \neq 0$ explicitly in Eqn.~\ref{TauKTfixedline} since this ``dangerously irrelevant'' coupling controls the long-distance correlations of $\tau_{\vec{r}}$ along this fixed-line.
Indeed, the nonzero value
of  $c_{\theta \tau}$ in ${\cal F}_{\tau {\rm KT}}$ causes $\tau_{\vec{r}}$ 
to inherit the power-law correlations of $\cos(3 \theta_{\vec{r}})$ for
all $T \in (T_{c1}, T_{c2})$: $\langle \tau_{\vec{r}} \tau_0\rangle \sim \langle e^{3i(\theta_{\vec{r}}-\theta_{0})}\rangle \sim 1/r^{9g(T)}$. Ferromagnetic couplings between the Ising
spins are not explicitly included in ${\cal F}_{\tau {\rm KT}}/T$ since the Ising bond-energy $E_{\langle \vec{r}_1 \vec{r}_2\rangle} \equiv \tau_{\vec{r}_1} \tau_{\vec{r}_2}$ has rapidly decaying
correlations $\langle E_{\langle \vec{r}_1 \vec{r}_2 \rangle} E_{\langle \vec{r}_3 \vec{r}_4\rangle} \rangle \sim 1/r^{36g}$ ($r$ is the distance between bonds $\langle \vec{r}_1 \vec{r}_2\rangle$ and $\langle \vec{r}_3 \vec{r}_4 \rangle$) that render these couplings irrelevant along this fixed line. Just below $g=1/9$ ({\em i.e.} for $T<T_{c1}$), the ferromagnetic couplings between the $\tau_{\vec{r}}$, and the six-fold anisotropy term $\cos(6\theta_{\vec{r}})$, both become relevant. This signals the
onset of six-fold symmetry-breaking long-range order in $H_{\rm eff}$.

{\bf Singular susceptibility:} For $\eta(T) < \frac{2}{9}$ in this power-law ordered intermediate phase of $H_{\rm eff}$, the foregoing
implies that power-law correlations of $\tau$ decay slowly enough that they lead to a {\em divergent} contribution $\chi_{\rm sing.} \sim L^{2-9\eta}$ to the finite-size susceptibility $\chi_L$ of an $L \times L$ system at $h=0$. This implies $\chi_L(T) = \chi_{\rm reg}(T) + b(T)L^{2-9\eta(T)}$ for $\eta(T) \in (\frac{1}{9},\frac{2}{9})$. When an external field $h$ is turned on in this regime, it  perturbs ${\cal F}_{\tau {\rm KT}}$
with a three-fold symmetric perturbation $J_{\theta \tau} \chi_{\rm reg} h \cos(3\theta_{\vec{r}})$.
This drives
$H_{\rm eff}$ to a long-range ordered state with correlation
length $\xi(h) \sim |h|^{-1/\lambda_3(g)}$, where $\lambda_3(g) = 2-9g/2$. Beyond this correlation-length scale,
$H_{\rm eff}$ resembles a three-state Potts model in its ordered state~\cite{Jose}.
Therefore, for small non-zero $h$,  $\chi_{\rm sing.}$ will be cut off at length-scales of order this correlation length $\xi(h)$, giving rise to a thermodynamic susceptibility
that scales as $(\xi(h))^{2-9\eta(T)}$ at small $h$. For
the thermodynamic easy-axis susceptibility of the microscopic easy-axis antiferromagnet, 
the foregoing analysis thus predicts
\begin{equation}
\chi(B) \sim |B|^{-\frac{4-18\eta(T)}{4-9\eta(T)}}
\end{equation}
at small $|B|$ for $\eta(T) \in (\frac{1}{9},\frac{2}{9})$.
This prediction identifies an experimentally useful signature of two-step melting of
either type (ferrimagnetic or antiferromagnetic) of three-sublattice order in triangular and
Kagome lattice easy-axis magnets. In particular, it applies to the $S=1$ triangular lattice Heisenberg antiferromagnet with strong single-ion anisotropy~\cite{Damle_Senthil,Heidarian_Damle}, and to the triangular lattice Ising antiferromagnet with further-neighbour couplings~\cite{Landau}.
It would therefore be interesting to identify quasi two-dimensional magnets in the Ca$_3$Co$_2$O$_6$ family~\cite{Paddison_etal,Stitzer_Darriet_Loye} (with an angular momentum $J=1$ ion at one Co site and a nonmagnetic ion at the other) which could provide experimental
realizations of the former.  It would also be interesting to identify new combinations of 
substrate and adsorbate for which monolayer densities closer to half-filling (than hitherto achievable~\cite{Bretz_etal,Bretz,Horn_Birgeneau_Heiney_Hammonds,Vilches,Suter_Colella_Gangwar,Feng_Chan,Wiechert}), corresponding to $B \ll 1$ in the latter,  could be reached for monolayers of adsorbed gases on triangular substrates.

{$\mathbf{T_{\tau} \gtrsim T_{\theta 2}}${\bf:}} In this case, $H_{\rm eff}$ develops
long-range order in $\tau$ via a transition in the Ising universality class at $T_{cI}$ (Fig~\ref{Fig2} A),
with the value of $T_{cI}$ set by $T_{\tau}$ (with deviations of order $J_{\theta \tau}$). For $T< T_{cI}$,
the spontaneous magnetization $\langle \tau \rangle$ perturbs $H_{\rm xy}$ with the three-fold field $h_{3{\rm eff}}$. Lowering the temperature below $T_{cI}$
along path $L$ in the phase diagram of $H_{\rm eff}$ (Fig.~\ref{Fig2} A)
therefore corresponds to moving along the eponymous path $L$ in the known~\cite{Jose} phase diagram of $H_{\rm xy}+ h_3 \sum_{\vec{r}} \cos(3 \theta_{\vec{r}})$ (Fig.~\ref{Fig2} B). This key observation immediately leads to two conclusions: First, $H_{\rm eff}$
must develop long-range order in $\theta$ at a {\em lower} temperature $T_{cP} < T_{cI}$ via a three-state Potts transition (Fig.~\ref{Fig2} A). 
Second, these Ising and three-state Potts transition lines ($T_{cI}$ and $T_{cP}$) must meet the phase boundaries of the power-law ordered phase ($T_{c2}$ and $T_{c1}$) at a single multicritical point ${\mathcal M}$ (Fig.~\ref{Fig2} A). 

{\bf Multicritical point:}   The fixed point theory $F_{{\mathcal M}}$ that controls
long-distance properties of ${\cal M}$ can be reached from the $c=3/2$
theory $F_{1/2}+ F_{\rm KT}$ (with $g=1/4$) by turning on the relevant perturbation $J_{\theta \tau}$. The $c$-theorem ~\cite{CardyBook} therefore
predicts that the central charge of $F_{{\mathcal M}}$ obeys $c_{{\mathcal M}} < \frac{3}{2}$.  Since $F_{\cal M}$ must have a relevant direction leading
from it to the $c=1$ theory $F_{\tau {\rm KT}}$, the $c$-theorem also predicts $ c_{{\mathcal M}} > 1$. At ${\mathcal M}$, the correlation functions $C_{\tau}(\vec{r}) = \langle \tau(\vec{r}) \tau(0)\rangle$  and $C_{p\theta} (\vec{r}) = \langle e^{i p \theta(\vec{r})} e^{-ip\theta(0)}\rangle $ ($p=1,2,3$) are expected to have the long distance forms: $C_{\tau}(\vec{r}) = 1/r^{\eta_{\tau}}$, $C_{p \theta}(\vec{r}) \sim 1/r^{\eta_{p\theta}}$ (with $\eta_{3\theta}= \eta_\tau$ on symmetry grounds). 
Setting $J_{\rm xy} = h_6 =1.0$, $J_{\theta \tau} = 0.25$ and parametrizing 
$J_{\rm Ising} = f_{xy}T_{\theta_1}/T_{\tau}$ and $T = f_{I}f_{xy}T_{\theta 1}$,
with $T_{\theta 1}=1.04$ and $T_{\tau}=3.6409$, I have performed extensive Monte-Carlo
simulations of $H_{\rm eff}$ to locate and study ${\mathcal M}$.
Fig~\ref{Fig3} displays power-law fits for the $L$ dependence of $C_{\tau}(\vec{r}_L)$ and $C_{p\theta}(\vec{r}_L)$ at separation $\vec{r}_L=\frac{L}{3}\hat{e}_x$ on periodic $L \times L$ triangular lattices at my best estimate
for ${\mathcal M}$, given by $[f_{\rm xy}^{{\mathcal M}},f_{I}^{{\mathcal M}}] \approx [1.5570(8), 1.0061(5)]$.
Such fits
yield the following estimates for multicritical exponents:
\begin{equation}
\! \! \!  \eta_{3\theta} = \eta_\tau \approx 0.201(20) \, ;  \eta_{\theta} \approx 0.258(5) \, ;  \eta_{2\theta} \approx 0.353(6) \, .
\end{equation}
This set of exponents is clearly different from the well-known exponents in the power-law ordered phase [$\eta_{p \theta} = p^2 \eta(T)$], or on the three-state Potts line [$\eta_{2\theta} = \eta_{\theta} = 4/15$] or the Ising line [$\eta_{3\theta} = \eta_{\tau} = 1/4$]. 
 
I close by noting an intriguing possibility: Since three-sublattice order melts via a three-state Potts transition in the Kagome Ising antiferromagnet with dipolar interactions~\cite{Moller_Moessner,Chern_etal,Chern_Tchernyshyov}, while the analogous short-ranged model with nearest and next-nearest neighbour exchange couplings exhibits two-step melting behaviour~\cite{Chern_Tchernyshyov,Wolf_Schotte,Takagi_Mekata}, it appears likely that such multicritical melting could be seen in artificial Kagome-ice systems if the strength of the first and second-neighbour exchange interactions could be increased relative to the long-range dipolar couplings (whose values are fixed by magnetostatics).

I thank F.~Alet, M.~Barma, D.~Dhar, and R.~Kaul for useful comments
on an earlier draft, G.~Mandal, S.~Minwalla, and S.~Trivedi for a survey of well-known $c>1$ conformal field theories, and R.~G.~Ghanshyam for help with figures.
The numerical work described here was made possible by the computational
resources of the Dept. of Theoretical Physics of the TIFR.
A major part of the analysis reported here was completed
while participating in the Program on Frustrated Magnetism and Quantum Spin Liquids
at KITP Santa Barbara, where this work was informed by useful discussions
with T.~Grover and a review of the artificial Kagome-ice literature by  R.~Moessner.
Participation in this program was made possible by partial support from National Science Foundation grant NSF PHY11-25915.

\end{document}